\documentclass[showpacs,preprintnumbers,amsmath,amssymb]{article}
\usepackage{cite}

\usepackage{epsfig}
\usepackage{graphicx}
\usepackage{dcolumn}
\usepackage{bm}
\usepackage[active]{srcltx}
\newcommand{\be}{\begin{equation}}

\newcommand{\ee}{\end{equation}}
\newcommand{\bea}{\begin{eqnarray}}
\newcommand{\eea}{\end{eqnarray}}

\usepackage{authblk}

\title{Causal information approach to partial conditioning in multivariate data sets}
\author[1]{D. Marinazzo\thanks{daniele.marinazzo@ugent.be}}
\author[2,3,4]{M. Pellicoro\thanks{mario.pellicoro@ba.infn.it}}
\author[2,3,4]{S. Stramaglia\thanks{sebastiano.stramaglia@ba.infn.it}}
\affil[1]{Department of Data Analysis, Faculty of Psychology and Pedagogical Sciences, University of Gent, B-9000 Gent, Belgium}
\affil[2]{Dipartimento  Interateneo di Fisica, Universit\`a di Bari, I-70126 Bari, Italy}
\affil[3]{TIRES-Center of Innovative Technologies for Signal Detection and Processing, Universit\`a di Bari,  Italy}
\affil[4]{I.N.F.N., Sezione di Bari, Italy}

\begin{document}

\date{\today}

\maketitle

\section*{Abstract}
When evaluating causal influence from one time series to another in a multivariate dataset it is necessary to take into account the conditioning effect of the other variables. In the presence of many variables, and possibly of a reduced number of samples, full conditioning can lead to computational and numerical problems.
In this paper we address the problem of partial conditioning to
a limited subset of variables, in the framework of information
theory. The proposed approach is tested on simulated datasets and on an example of intracranial EEG recording from an epileptic subject. We show that, in many instances, conditioning on a
small number of variables, chosen as the most informative ones for
the driver node, leads to results very close to those obtained with a fully
multivariate analysis, and even better in the presence of a small number of samples. This is particularly relevant when the
pattern of causalities is sparse.

\section*{Introduction}

Determining how the brain is connected is a crucial point in neuroscience.
To gain better understanding of which neurophysiological processes are linked to which brain mechanisms, structural connectivity in the brain can be complemented by the investigation of statistical dependencies between distant brain regions (functional connectivity), or of models aimed to elucidate drive-response relationships (effective connectivity).
 Advances in imaging techniques guarantee an immediate improvement in our knowledge of structural connectivity. A constant computational and modelling effort has to be done in order to optimize and adapt functional and effective connectivity to the qualitative and quantitative changes in data and physiological applications.  The paths of information flow throughout the brain can shed light on its functionality in health and pathology.
Every time that we record brain activity we can imagine that we are monitoring the activity at the nodes of a network.
This activity is dynamical and sometimes chaotic. Dynamical networks \cite{barabasi_linked_2002} model physical and biological
behaviour in many applications; also, synchronization in dynamical network is influenced by the topology
of the network itself\cite{boccaletti_synchronization_2006}. A great need
exists for the development of effective methods of inferring network
structure from time series data; a dynamic version of Bayesian
Networks has been proposed in \cite{gharhamani_1997}.  A method for
detecting the topology of dynamical networks, based on chaotic
synchronization, has been proposed in \cite{yu_estimating_2006}.

Granger causality has become the method of choice to determine
whether and how two time series exert causal influences on each
other \cite{hlavackova-schindler_causality_2007},\cite{bressler_wieneraGranger_2011}. This approach is based on prediction: if the
prediction error of the first time series is reduced by including
measurements from the second one in the linear regression model,
then the second time series is said to have a causal influence on
the first one. This frame has been used in many fields of science,
including neural systems \cite{kaminski_evaluating_2001},\cite{blinowska_Granger_2004},\cite{seth_causal_2005},\cite{roebroeck_mapping_2005}, reo-chaos \cite{ganapathy_Granger_2007} and
cardiovascular variability \cite{faes_assessment_2008}.

From the beginning \cite{Granger69},\cite{wiener_1956}, it has been
known that if two signals are influenced by third one that is not
included in the regressions, this leads to spurious causalities, so
an extension to the multivariate case is in order. The conditional
Granger causality analysis (CGCA) \cite{geweke_measures_1984} is
based on a straightforward expansion of the autoregressive model to
a general multivariate case including all measured variables. CGCA
has been proposed to correctly estimate coupling in multivariate
data sets
\cite{barrett_multivariate_2010},\cite{chen_frequency_2006},\cite{deshpande_multivariate_2009},\cite{zhou_analyzing_2009}.
Sometimes though, a fully multivariate approach can entrain problems
which can be purely computational but even conceptual: in presence
of redundant variables the application of the standard analysis
leads to under-estimation of causalities
\cite{angelini_redundant_2010}.

Several approaches have been proposed in order to reduce dimensionality in multivariate sets, relying on generalized variance \cite{barrett_multivariate_2010},
principal components analysis \cite{zhou_analyzing_2009} or Granger causality itself \cite{marinazzo_grouping_2010}.

In this paper we will address the problem of partial conditioning to
a limited subset of variables, in the framework of information
theory. Intuitively, one may expect that conditioning on a small
number of variables should be sufficient to remove indirect
interactions if the connectivity pattern is sparse. We will show
that this subgroup of variables might be chosen as the most
informative for the driver variable, and describe the application to
simulated examples and a real data set.
\section*{Materials and Methods}

We start by describing the connection between Granger causality and
information-theoretic approaches like the transfer entropy in
\cite{schreiber_measuring_2000}. Let $\{\xi_n\}_{n=1,.,N+m}$ be a time series that
may be approximated by a stationary Markov process of order $m$,
i.e.
$p(\xi_n|\xi_{n-1},\ldots,\xi_{n-m})=p(\xi_n|\xi_{n-1},\ldots,\xi_{n-m-1})$.
We will use the shorthand notation
$X_i=(\xi_{i},\ldots,\xi_{i+m-1})^\top$ and $x_i=\xi_{i+m}$, for
$i=1,\ldots,N$, and treat these quantities as $N$ realizations of
the stochastic variables $X$ and $x$. The minimizer of the risk
functional
\begin{equation}\label{risk}
\mbox{\cal{R}} \left[ f \right] = \int dX dx \left( x - f(X)
\right)^2 p(X, x)
\end{equation} represents the best estimate of $x$, given X, and
corresponds \cite{papoulis_1985} to the regression function $f^*(X)
= \int dx p(x | X )x$. Now, let $\{\eta_n\}_{n=1,.,N+m}$ be another
time series of simultaneously acquired quantities, and denote
$Y_i=(\eta_{i},\ldots,\eta_{i+m-1})^\top$. The best estimate of $x$,
given $X$ and $Y$, is now: $g^*(X,Y)=\int dx  p(x | X,Y )x$. If the
generalized Markov property holds, i.e. \be \label{genmarkov}p(x |
X,Y )=p(x | X ),\ee then $f^*(X)=g^*(X,Y)$ and the knowledge of $Y$
does not improve the prediction of $x$. Transfer entropy
\cite{schreiber_measuring_2000} is a measure of the violation of
\ref{genmarkov}: it follows that Granger causality implies non-zero
transfer entropy \cite{marinazzo_kernel_2008}. Under Gaussian
assumption it can be shown that Granger causality and transfer
entropy are entirely equivalent, and just differ for a factor two
\cite{barnett_Granger_2009}. The generalization of Granger causality
to a multivariate fashion, described in the following, allows the
analysis of dynamical networks \cite{Marinazzopre2008} and to
discern between direct and indirect interactions.

Let us consider $n$ time series $\{x_\alpha (t)\}_{\alpha
=1,\ldots,n}$; the state vectors are denoted
$$X_\alpha (t)= \left(x_\alpha (t-m),\ldots,x_\alpha (t-1)\right),$$
$m$ being the window length (the choice of $m$ can be done using the
standard cross-validation scheme). Let $\epsilon \left(x_\alpha
|{\bf X}\right)$ be the mean squared error prediction of $x_\alpha$
on the basis of all the vectors ${\bf X}$ (corresponding to linear
regression or non linear regression by the kernel approach described
in \cite{marinazzo_kernel_2008}). The multivariate Granger causality
index $c (\beta \to \alpha )$ is defined as follows: consider the
prediction of $x_\alpha$ on the basis of all the variables but
$X_\beta$ and the prediction of $x_\alpha$ using all the variables,
then the causality measures the variation of the error in the two
conditions, i.e.
\begin{equation}\label{delta}
c(\beta \to \alpha )=\log{\epsilon \left(x_\alpha |{\bf X}\setminus
X_\beta\right)\over \epsilon \left(x_\alpha |{\bf X}\right)}.
\end{equation}
Note that in \cite{marinazzo_kernel_2008} a different definition of
causality has been used,
\begin{equation}\label{delta1} \delta (\beta \to \alpha )={\epsilon
\left(x_\alpha |{\bf X}\setminus X_\beta\right)-\epsilon
\left(x_\alpha |{\bf X}\right)\over \epsilon \left(x_\alpha |{\bf
X}\setminus X_\beta\right)};
\end{equation} The two definitions are clearly related by a monotonic
transformation:
\begin{equation}\label{trans}
c(\beta \to \alpha )=-\log{\left[1-\delta (\beta \to \alpha
)\right]}.
\end{equation}
Here we first evaluate the causality $\delta (\beta \to \alpha )$
using the selection of significative eigenvalues described in
\cite{Marinazzopre2008} to address the problem of over-fitting in
(\ref{delta1}); then we use (\ref{trans}) and express our results in
terms of $c(\beta \to \alpha )$, because it is with this definition
that causality is twice the transfer entropy, equal to $I\{x_\alpha
;X_\beta |{\bf X}\setminus X_\beta \}$, in the Gaussian case
\cite{barnett_Granger_2009}.

Turning now to the central point of this paper, we address the
problem of coping with a large number of variables, when the
application of multivariate Granger causality may be questionable or
even unfeasible, whilst bivariate causality would detect also
indirect causalities. Here we show that conditioning on a small
number of variables, chosen as the most informative for the
candidate driver variable, is sufficient to remove indirect
interactions for sparse connectivity patterns. Conditioning on a
large number of variables requires an high number of samples in
order to get reliable results. Reducing the number of variables,
that one has to condition over, would thus provide better results
for small data-sets. In the general formulation of Granger causality, one has no way to choose this reduced set of variables; on the other hand, in the framework of information theory, it is possible to individuate the most informative variables one by one. Once that it has been demonstrated \cite{barnett_Granger_2009} that Granger causality is
equivalent to the  information flow between Gaussian variables, partial conditioning becomes possible for Granger causality estimation;  to our knowledge this is the first time that such approach is proposed.

Concretely, let us consider the causality $\beta \to \alpha$; we fix
the number of variables, to be used for conditioning, equal to
$n_d$. We denote ${\bf Z}=\left(X_{i_1},\ldots,X_{i_{n_d}}\right)$
the set of the $n_d$ variables, in ${\bf X}\setminus X_\beta$, most
informative for $X_\beta$. In other words, {\bf Z} maximizes the
mutual information $I\{X_\beta ; {\bf Z}\}$ among all the subsets
${\bf Z}$ of $n_d$ variables. Then, we evaluate the causality
\begin{equation}\label{delta2}
c (\beta \to \alpha )=\log{\epsilon \left(x_\alpha |{\bf
Z}\right)\over \epsilon \left(x_\alpha |{\bf Z}\cup X_\beta\right)}.
\end{equation}
Under the Gaussian assumption, the mutual information $I\{X_\beta ;
{\bf Z}\}$ can be easily evaluated, see \cite{barnett_Granger_2009}.
Moreover, instead of searching among all the subsets of $n_d$
variables, we adopt the following approximate strategy. Firstly the
mutual information of the driver variable, and each of the other
variables, is estimated, in order to choose the first variable of
the subset. The second variable of the subsets is selected among the
remaining ones, as those that, jointly with the previously chosen
variable, maximizes the mutual information with the driver variable.
Then, one keeps adding the rest of the variables by iterating this
procedure. Calling ${\bf Z }_{k-1}$ the selected set of $k-1$
variables, the set ${\bf Z }_{k}$ is obtained adding , to ${\bf Z
}_{k-1}$, the variable, among the remaining ones,  with greatest
information gain. This is repeated until $n_d$ variables are
selected. This greedy algorithm, for the selection of relevant
variables, is expected to give good results under the assumption of
sparseness of the connectivity.

\section*{Results and Discussion}
\subsection*{Simulated data}
\begin{figure}[ht]
\includegraphics[width=10cm]{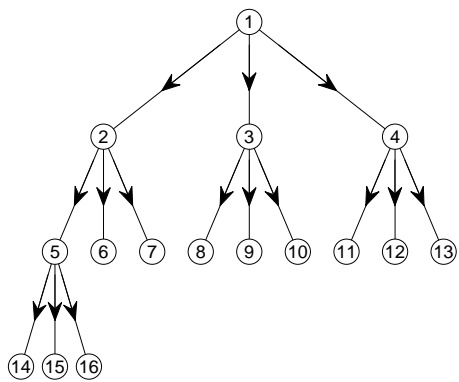}
\caption{{\rm  A directed rooted tree of 16 nodes.
\label{fig1}}}\end{figure}

\begin{figure}[ht]
\includegraphics[width=10cm]{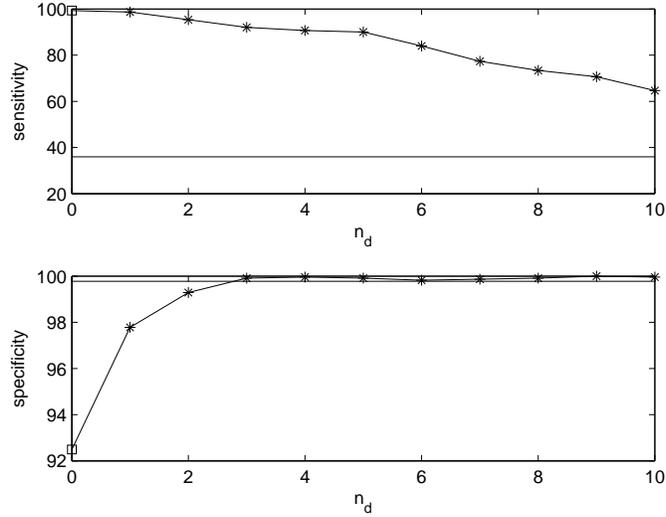}
\caption{{\rm The sensitivity (top) and the specificity (bottom) are
plotted versus $n_d$, the number of variables selected for
conditioning, for the first example, the rooted tree.  The number of
samples $N$ is 100 and the order is $m=2$; similar results are
obtained varying $m$. The results are averaged over 100 realizations
of the linear dynamical system described in the text. The empty
square, in correspondence to $n_d =0$, is the result from the
bivariate analysis. The horizontal line is the outcome from
multivariate analysis, where all variables are used for
conditioning. \label{fig2}}}\end{figure}

Let us consider linear dynamical systems on a lattice of $n$ nodes,
with equations, for $i=1,\ldots,n$:
\begin{equation}
x_{i,t}=\sum_{j=1}^n a_{ij} x_{j,t-1}+s \tau_{i,t}, \label{lds}
\end{equation}
where $a$'s are the couplings, $s$ is the strength of the noise and
$\tau$'s are unit variance i.i.d. Gaussian noise terms. The level of noise determines the minimal amount of samples needed to assess that the structures recovered by the proposed approachare genuine and are not due to randomness, as it happens for the standard Granger causality (see discussions in \cite{marinazzo_kernel_2008} and \cite{Marinazzopre2008}); in particular noise should not be too high to obscure deterministic effects.  Firstly we
consider a directed tree of 16 nodes depicted in figure
(\ref{fig1}); we set $a_{ij}$ equal to 0.9 for each directed link of
the graph thus obtained, and zero otherwise. We set $s=0.1$. In
figure (\ref{fig2}) we show the application of the proposed
methodology to data sets generated by eqs. (\ref{lds}), 100 samples
long, in terms of quality of the retrieved network, expressed in
terms of sensitivity (the percentage of existing links that are
detected) and specificity (the percentage of missing links that are
correctly recognized as non existing). The bivariate analysis
provides 100$\%$ sensitivity and 92$\%$ specificity.  However
conditioning on a few variables is sufficient to put in evidence
just the direct causalities while still obtaining high values of
sensitivity. The full multivariate analysis (obtained as $n_d$ tends
to 16) gives here a rather low sensitivity, due to the low number of
samples. This is a clear example where conditioning on a small
number of variables is better than conditioning on all the variables
at hand.

\begin{figure}[ht]
\includegraphics[width=10cm]{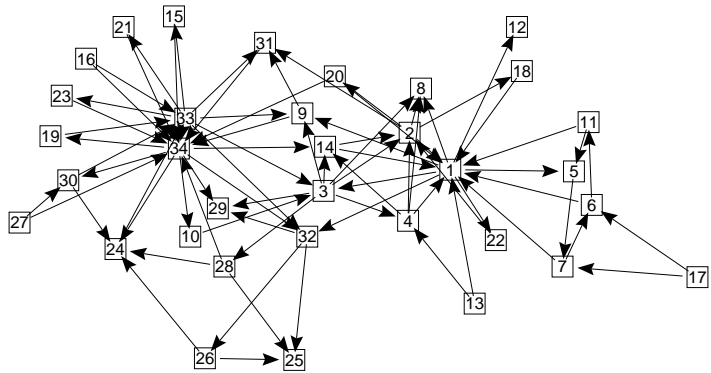}
\caption{{\rm  The directed network of 34 nodes obtained assigning
randomly a direction to links of the Zachary network.
\label{fig3}}}\end{figure}

As another example, we now fix $n=34$ and construct couplings in
terms of the well known Zachary data set \cite{zachary_1977}, an
undirected network of $34$ nodes. We assign a direction to each
link, with equal probability, and set $a_{ij}$ equal to 0.015, for
each link of the directed graph thus obtained, and zero otherwise.
The noise level is set $s=0.5$. The network is displayed in figure
(\ref{fig3}): the goal is again to estimate this directed network
from the measurements of time series on nodes.

\begin{figure}[ht]
\includegraphics[width=10cm]{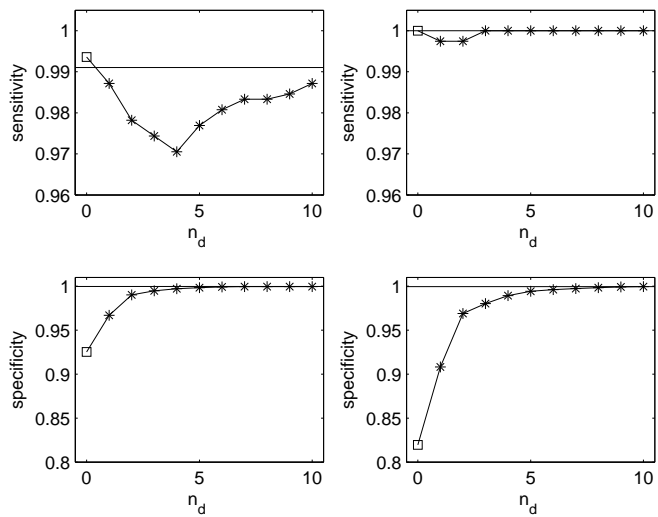}
\caption{{\rm  Sensitivity and specificity are plotted versus $n_d$,
the number of variables selected for conditioning, for two values of
two values of the number of samples $N$, 500 (left) and 1000
(right). The order is $m=2$, similar results are obtained varying
$m$. The results are averaged over 100 realizations of the linear
dynamical system described in the text. The empty square, in
correspondence to $n_d =0$, is the result from the bivariate
analysis. The horizontal line is the outcome from multivariate
analysis, where all variables are used for conditioning.
\label{fig4}}}\end{figure}

\begin{figure}[ht]
\includegraphics[width=10cm]{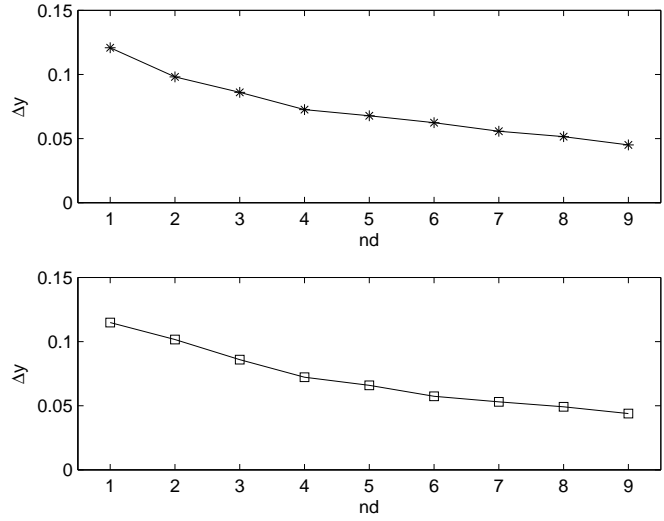}
\caption{{\rm  The mutual information gain, when the $(n_d +1)$-th
variable is included, is plotted versus $n_d$ for two values of the
of the number of samples $N$, 500 (top) and 1000 (bottom). The order
is $m=2$. The information gain is averaged over all the variables.
\label{fig5}}}\end{figure}

In figure (\ref{fig4}) we show the application of the proposed
methodology to data sets generated by eqs. (\ref{lds}), in terms of
sensitivity and specificity, for different numbers of samples. The bivariate analysis detects several
false interactions, however conditioning on a few variables is
sufficient to put in evidence just the direct causalities. Due to
the sparseness of the underlying graph, we  get a result which is
very close to the one by the full multivariate analysis; the
multivariate analysis here recovers the true network, indeed the
number of samples is sufficiently high. In figure (\ref{fig5}),
concerning the stage of selection of variables upon which
conditioning, we plot the mutual information gain as a function of
the number of variables included $n_d$: it decreases as $n_d$
increases.

\clearpage

%

\clearpage
\subsection*{EEG epilepsy data}
\begin{figure}[ht]
\includegraphics[width=10cm]{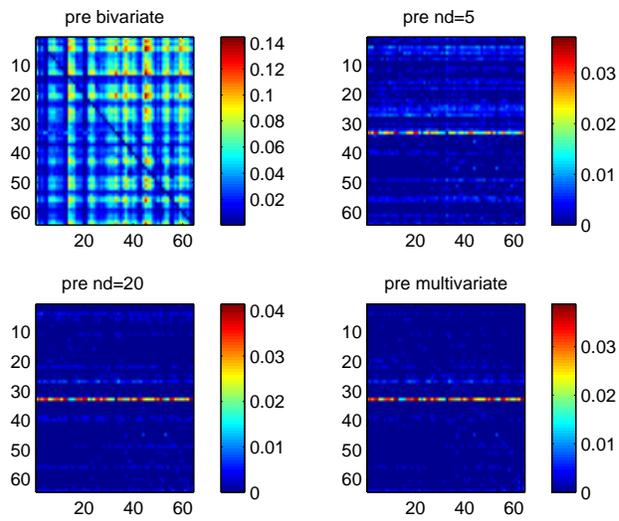}
\caption{{\rm  The causality analysis of the preictal period. The
causality c($i\to j$) corresponds to the row i and the column j. The
order is chosen $m=6$ according to the AIC criterion. Top left:
bivariate analysis. Top right: our approach with $n_d =5$
conditioning variables. Bottom left: our approach with $n_d =20$
conditioning variables. Bottom right: the multivariate analysis.
\label{fig6}}}\end{figure}

\begin{figure}[ht]
\includegraphics[width=10cm]{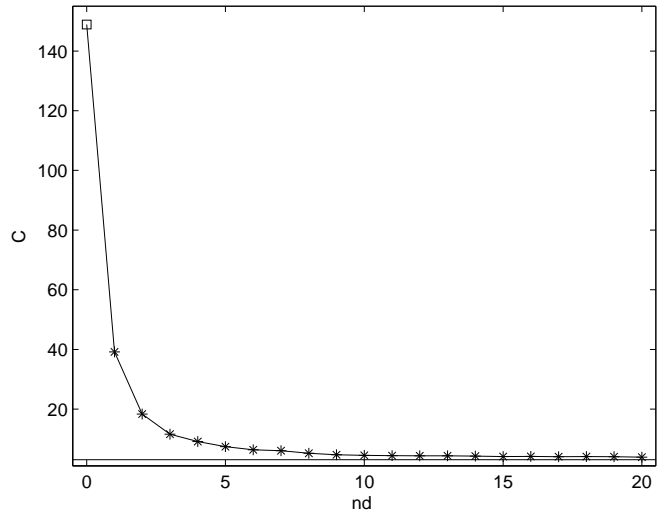}
\caption{{\rm Concerning the preictal period, the sum of all
causalities is plotted versus the number of conditioning variables
\label{fig7}}}\end{figure}

We consider now a real data set from an $8\times 8$ electrode grid
that was implanted in the cortical surface of the brain of a patient
with epilepsy \cite{kramer}. We consider two 10-seconds intervals prior to and immediately after the onset of a
seizure, called respectively the preictal period and the ictal period. In
figure (\ref{fig6}) we show the application of our approach to the
preictal period; we used the linear causality. The bivariate
approach detects many causalities between the electrodes; most of
them, however, are indirect. According to the multivariate analysis
there is just one electrode which is observed to influence the
others, even in the multivariate analysis: this electrode
corresponds to a localized source of information and could indicate
a putative epileptic focus. In (\ref{fig6}) it is shown that
conditioning on $n_d =5$ or $n_d =20$ variables provides the same
pattern corresponding to the multivariate analysis, which thus
appears to be robust.  These results suggest that the effective
connectivity is sparse in the preictal period. As a further
confirmation, in (\ref{fig7}) we plot the sum of all causalities
detected as a function of the number of conditioning variables, for
the preictal period; a plateau is reached already for small values
of $n_d$.

\begin{figure}[ht]
\includegraphics[width=10cm]{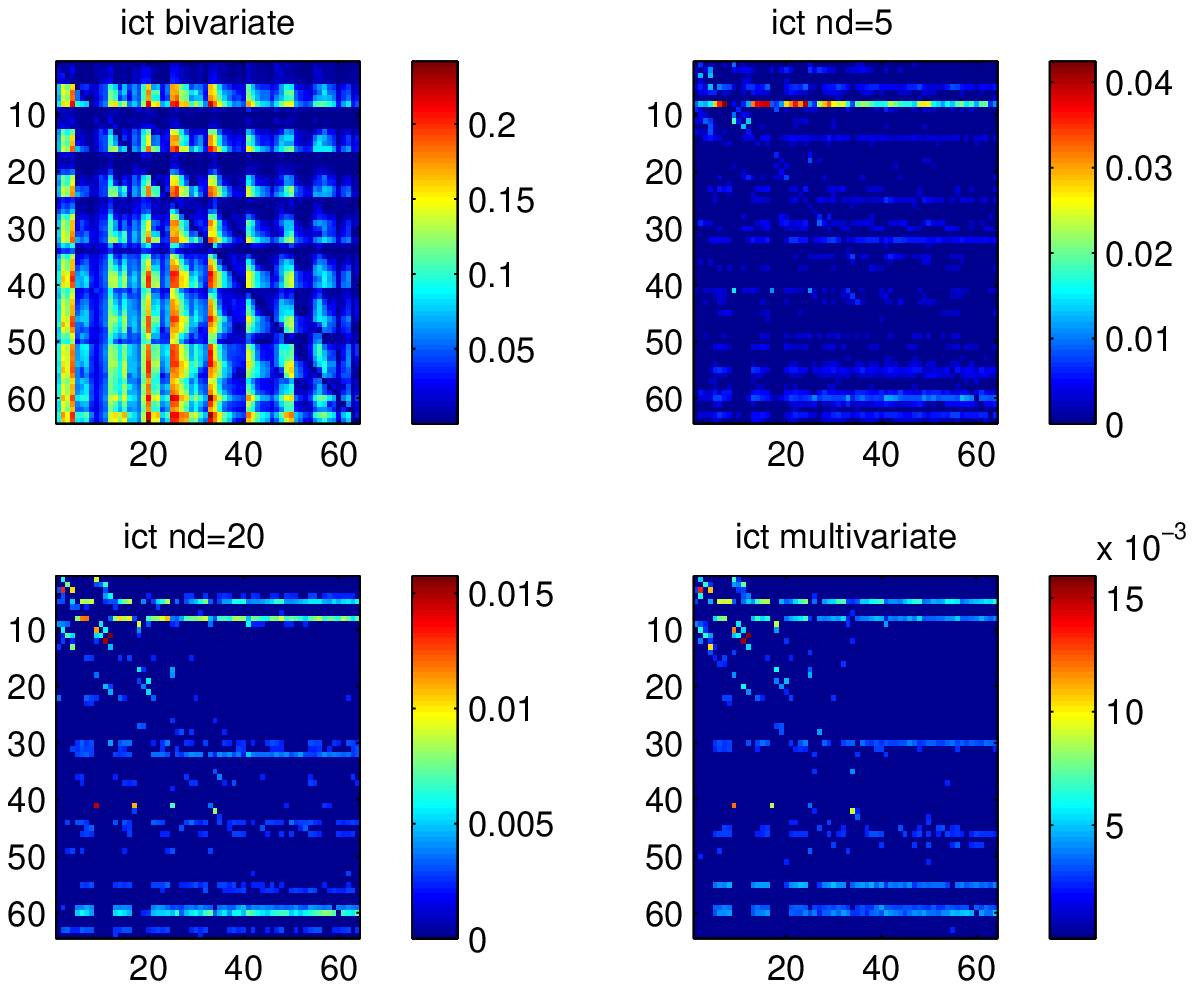}
\caption{{\rm  The sum of outgoing causality from each electrode in
the EEG application, ictal period. Top left: bivariate analysis.
Top right: our approach with $n_d =5$ conditioning variables. Bottom
left: our approach with $n_d =20$ conditioning variables. Bottom
right: the multivariate analysis.\label{fig8}}}\end{figure}

\begin{figure}[ht]
\includegraphics[width=10cm]{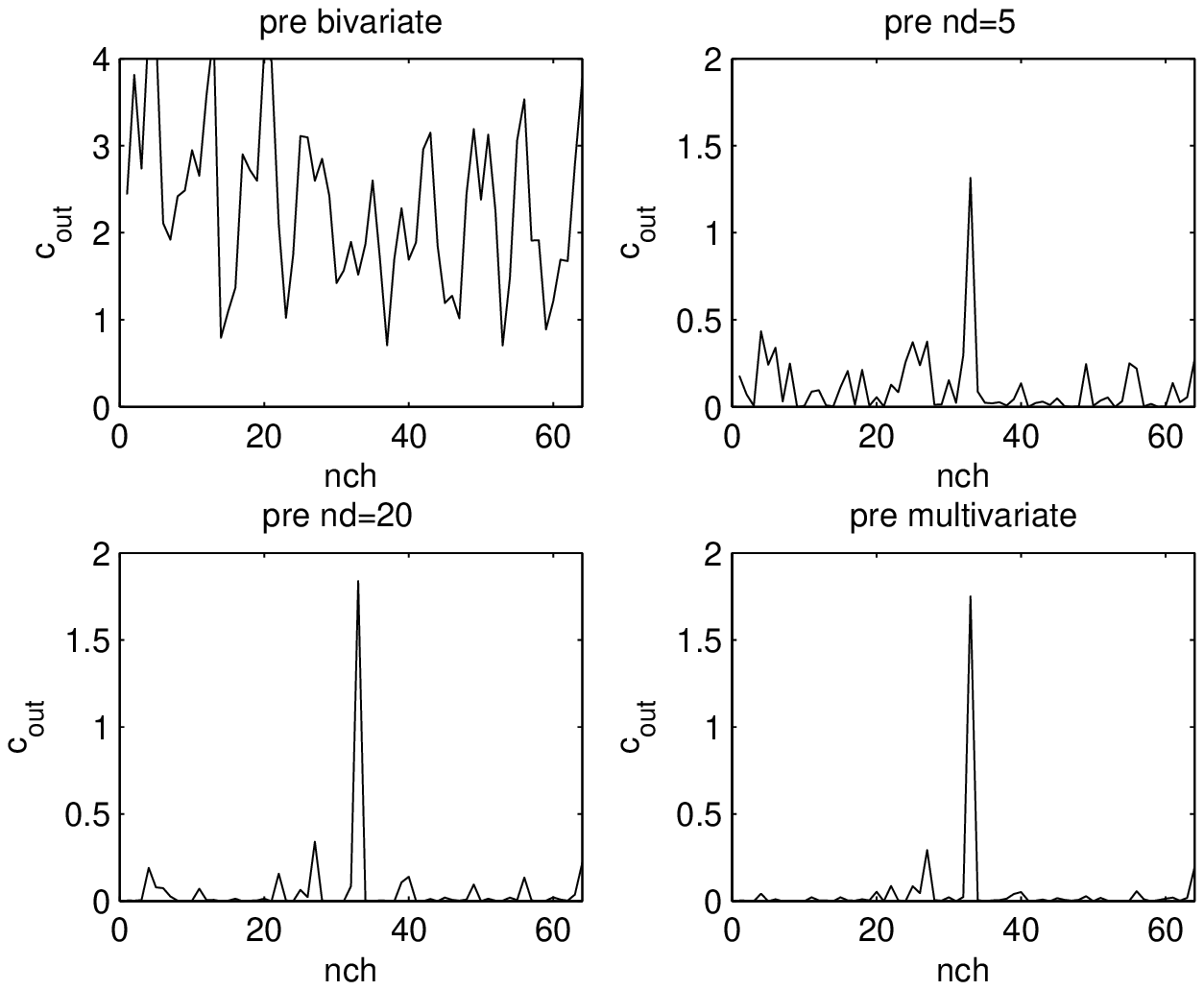}
\caption{{\rm  The sum of outgoing causality from each electrode in
the EEG application, preictal period. Top left: bivariate analysis. Top
right: our approach with $n_d =5$ conditioning variables. Bottom
left: our approach with $n_d =20$ conditioning variables. Bottom
right: the multivariate analysis. \label{fig9}}}\end{figure}

\begin{figure}[ht]
\includegraphics[width=10cm]{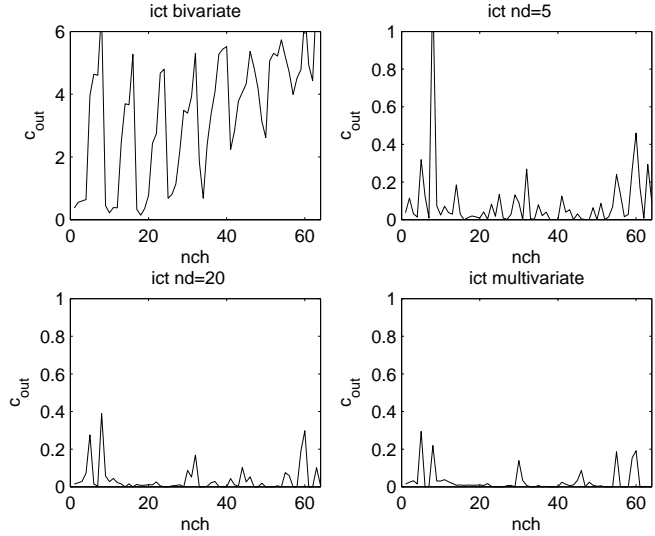}
\caption{{\rm  The causality analysis of the ictal period. The
causality c($i\to j$) corresponds to the row i and the column j. The
order is chosen $m=6$ according to the AIC criterion. Top left:
bivariate analysis. Top right: our approach with $n_d =5$
conditioning variables. Bottom left: our approach with $n_d =20$
conditioning variables. Bottom right: the multivariate analysis.
\label{fig10}}}\end{figure}

In (\ref{fig8}) the same analysis is shown w.r.t. the ictal period:
in this case conditioning on $n_d =5$ or $n_d =20$ variables does
not reproduce the pattern obtained with the multivariate approach. The
lack of robustness of the causality pattern w.r.t. $n_d$ seems to
suggest that the effective connectivity pattern, during the crisis,
is not sparse. In (\ref{fig9}) and (\ref{fig10}) we show, for each
electrode and for the preictal and ictal periods respectively, the
total outgoing causality (obtained as the sum of the causalities on
all the other variables). These pictures confirm the discussion
above: looking at how the causality changes with $n_d$ may provide
information about the sparseness of the effective connectivity.

 \clearpage
 \clearpage
\clearpage
\clearpage

\section*{Conclusions}
We have addressed the problem of partial conditioning to a limited
subset of variables while estimating causal connectivity, as an
alternative to full conditioning, which can lead to computational
and numerical problems. Analyzing  simulated examples and a real
data-set, we have shown that conditioning on a small number of
variables, chosen as the most informative ones for the driver node,
leads to results very close to those obtained with a fully
multivariate analysis, and even better in the presence of a small
number of samples, especially when the pattern of causalities is
sparse. Moreover, looking at how causality changes with the number
of conditioning variables provides information about the sparseness
of the connectivity.


\bibliography{mybib}
\bibliographystyle{abbrv}

%
%

\end{document}